
%
%
%
%
\input harvmac.tex

\Title{\vbox{\baselineskip12pt
\hbox{LAVAL-PHY-20/95}}}
{\vbox {\centerline{ Quantum chains with a Catalan tree pattern}
\bigskip
\centerline{ of conserved charges:}
\bigskip
\centerline{the  $\Delta = -1$ XXZ model and the isotropic octonionic chain}
}}
\bigskip
\bigskip

\centerline{M. P. Grabowski$^a$ and P. Mathieu$^b$\foot{Work
supported by NSERC (Canada). } }

\smallskip

\centerline{ \it D\'epartement de
physique, Universit\'e Laval, Qu\'ebec, Canada G1K 7P4} \vskip.05in
\centerline{$^a$ e-mail: mgrabows@phy.ulaval.ca}
\centerline{$^b$ e-mail: pmathieu@phy.ulaval.ca}

\vskip .2in
\centerline{\bf Abstract}
\smallskip
A class of quantum chains possessing a family of local conserved charges with a
Catalan tree pattern is studied. Recently, we have identified
such a structure
in the integrable $SU(N)$-invariant chains. In the present work
we find sufficient conditions for the existence of a family
of charges with this structure in terms of the underlying algebra.
Two additional systems with a Catalan tree structure of conserved
charges are found. One is the spin 1/2 XXZ model with
$\Delta=-1$. The other is a new  octonionic isotropic chain,
generalizing the Heisenberg model. This system provides an interesting
example of an infinite family of noncommuting local conserved quantities.
\smallskip
\smallskip
\bigskip
\noindent
 \Date{01/95}

\newcount\eqnum \eqnum=1
\def\eq{
\eqno(\secsym\the\meqno)
\global\advance\meqno by1
 }
\def\eqlabel#1{
{\xdef#1{\secsym\the\meqno}}
\eq
}

\newwrite\refs
\def\startreferences{
 \immediate\openout\refs=references
 \immediate\write\refs{\baselineskip=14pt \parindent=16pt \parskip=2pt}
}
\startreferences

\refno=0
\def\aref#1{\global\advance\refno by1
 \immediate\write\refs{\noexpand\item{\the\refno.}#1\hfil\par}}
\def\ref#1{\aref{#1}\the\refno}
\def\refname#1{\xdef#1{\the\refno}}

\def\s{\sigma}

\def\la{\lambda}

\def\ep{\epsilon}

\def\ra{\rightarrow}

\let\n=\noindent
\def\frac#1#2{{\textstyle{#1\over #2}}}

\font\smallcapfont=cmr9
\def\sc#1{{\smallcapfont\uppercase{#1}}}

\def \calC{{\cal {C}}}
\def \calB{{\cal {B}}}
\def \calA{{\cal {A}}}

\def\text#1{\quad\hbox{#1}\quad}


\def\ubrackfill#1{$\mathsurround=0pt
	\kern2.5pt\vrule depth#1\leaders\hrule\hfill\vrule depth#1\kern2.5pt$}
\def\contract#1{\mathop{\vbox{\ialign{##\crcr\noalign{\kern3pt}
	\ubrackfill{4pt}\crcr\noalign{\kern3pt\nointerlineskip}
	$\hfil\displaystyle{#1}\hfil$\crcr}}}\limits
}

\def\ubrack#1{$\mathsurround=0pt
	\vrule depth#1\leaders\hrule\hfill\vrule depth#1$}
\def\dbrack#1{$\mathsurround=0pt
	\vrule height#1\leaders\hrule\hfill\vrule height#1$}
\def\ucontract#1#2{\mathop{\vbox{\ialign{##\crcr\noalign{\kern 4pt}
	\ubrack{#2}\crcr\noalign{\kern 4pt\nointerlineskip}
	$\hskip #1\relax$\crcr}}}\limits
}
\def\dcontract#1#2{\mathop{\vbox{\ialign{##\crcr
	$\hskip #1\relax$\crcr\noalign{\kern0pt}
	\dbrack{#2}\crcr\noalign{\kern0pt\nointerlineskip}
	}}}\limits
}

\def\ucont#1#2#3{^{\kern-#3\ucontract{#1}{#2}\kern #3\kern-#1}}
\def\dcont#1#2#3{_{\kern-#3\dcontract{#1}{#2}\kern #3\kern-#1}}

\def \sumL{{\sum_{j\in\Lambda}}}
\def \L{{\Lambda}}
\def \S{{\bf {\sigma} }}
\def \u{{\bf {u} }}
\def \a{{\la_x}}
\def \b{{\la_y}}
\def \c {{\la_z}}
\def \aone{{\kappa_1}}

\def \atwo{{\kappa_2}}
\def \btwo{{b_2}}
\def \athree{{\kappa_3}}
\def \bthree{{b_3}}

\font\tenmib=cmmib10
\font\sevenmib=cmmib10 at 7pt
\font\fivemib=cmmib10 at 5pt
\newfam\mibfam 

\textfont\mibfam=\tenmib
\scriptfont\mibfam=\sevenmib
\scriptscriptfont\mibfam=\fivemib
\mathchardef\alphaB="080B
\mathchardef\betaB="080C
\mathchardef\gammaB="080D
\mathchardef\deltaB="080E
\mathchardef\epsilonB="080F
\mathchardef\zetaB="0810
\mathchardef\etaB="0811
\mathchardef\thetaB="0812
\mathchardef\iotaB="0813
\mathchardef\kappaB="0814
\mathchardef\lambdaB="0815
\mathchardef\muB="0816
\mathchardef\nuB="0817
\mathchardef\xiB="0818
\mathchardef\piB="0819
\mathchardef\rhoB="081A
\mathchardef\sigmaB="081B
\mathchardef\tauB="081C
\mathchardef\upsilonB="081D
\mathchardef\phiB="081E
\mathchardef\chiB="081F
\mathchardef\psiB="0820
\mathchardef\omegaB="0821
\mathchardef\varepsilonB="0822
\mathchardef\varthetaB="0823
\mathchardef\varpiB="0824
\mathchardef\varrhoB="0825
\mathchardef\varsigmaB="0826
\mathchardef\varphiB="0827

\def \calC{{\cal {C}}}

\def \calA{{\cal {A}}}
\def \calB{{\cal {B}}}

\def \mult {{\cdot}}

\def \r#1{{}}
\def \V{{\bf {v}}}

\def \a{{\la_x}}
\def \b{{\la_y}}
\def \c {{\la_z}}
\def \S{\sigmaB }

\def \Sh {{\hat \S}}
\def \sh {{\hat \s}}
\def \St {{\tilde \S}}
\def \st {{\tilde \s}}
\def \ft {{\tilde f}}


\newsec{Introduction}

A remarkably simple explicit expression for all the local conserved charges of
the  periodic or infinite XXX model (spin 1/2 quantum chain) has been found
recently [\ref{Grabowski, M.P., Mathieu,  P.: Quantum integrals of motion for
the Heisenberg spin chain.
Mod. Phys. Lett. {\bf 9A}, 2197-2206 (1994) (hep-th
9403149)}\refname\GMa,\ref{Grabowski, M.P., Mathieu, P.:
{The structure of
the conservation laws in integrable spin chains with short-range interactions}.
Preprint Laval-PHY-21/94 (hep-th 9411045)}\refname\GMb] (see also
[\ref{ Anshelevich, V.V.: First integrals and stationary states of the
Heisenberg quantum spin dynamics.  Theor. Math. Phys.
{\bf 43}, 350-353 (1980)}\refname\Ans] for an earlier presentation of an
equivalent
result for the infinite chain). In an appropriate basis, each charge is
described in terms a Catalan tree
pattern.  A direct argument, based on three simple identities, has been devised
to prove that these integrals commute with the hamiltonian and among
themselves.  This provides thus an alternative and new proof of the
integrability
of the XXX model.  But this in itself is not very interesting given that the
integrability of this model,
which is probably the most intensively studied
quantum integrable system, has
already been proved in diverse ways. The interest, of course, lies in the
possibility of generalizing the argument in view of discovering new integrable
systems or even unraveling the structure of the conserved charges of other
known
integrable systems.

Such an
extension has already been found in [\GMb], where it is shown that the XXX
pattern
is in fact common to all isotropic $su(N)$ quantum models formulated in their
fundamental representation. The
proof in that case simply boils down to showing that the three identities
alluded to above hold true
in this case, when the Pauli matrices
are replaced by the Gell-Mann matrices and
the $su(2)$ structure constants by their $su(N)$ counterparts. The
fundamental representation is singled out from the requirement that the
anticommutator of the generators should close in the algebra, a
necessary condition for the validity of two of the identities.

In view of enlarging the potential applications of this direct method, we
consider here two sorts of generalization of the three identities at the core
of the algebraic direct method. As a first step, we reformulate the model in
terms of an unspecified algebra, and introduce as many free parameters as
possible in the three identities.  When the normalization of the
generators is fixed, two
such free parameters can be introduced. Once the algebra will be specified,
these
parameters will automatically be determined from the expression for the
contraction of two (resp. three) indices in the product of two (resp. three)
structure constants.  Notice that the hamiltonian is the sum of the
nearest-neighbor
bilinear in the algebra generators and it does not contain any parameter, i.e.
it
is isotropic.  One then constructs the generalization of the XXX charges, with
$su(2)$ quantities replaced by their analogues in the yet unspecified algebra,
and
see under which conditions these charges commute with the hamiltonian.  These
conditions turn out to be simply a linear relation between our two free
parameters. Thus we end up with a one parameter set of identities, sufficient
to
guarantee the existence of an infinite number of conservation laws
for an infinite chain.
By fixing the algebra, which amounts to
fixing the two parameters, we can thus readily check
whether the linear relation between the
two now-determined parameters is satisfied.
If it is, then not only the
existence of an infinite number of conserved charges is automatically proved,
but as a bonus, we have an explicit
expression for them.
By construction, these charges have a Catalan tree pattern.
Moreover, for a particular value of the free parameter, our construction
ensures automatically mutual commutativity of all the charges, hence
demonstrating  the integrability of the model.

Here we present a new quantum chain, for which
the existence of an infinite family of conservation laws can be
proved exactly in this way. This is the isotropic octonionic chain. Since
the $su(2)$ XXX model is actually a quaternionic chain,
this generalization looks, in retrospect, rather natural.

Another direction where a generalization can be contemplated is for anisotropic
models.  Considering for concreteness the XYZ model, we can determine by a
systematic procedure how the building blocks of the XXX charges (i.e. the
Catalan
tree vertices) would have to be modified to take into account the
anisotropy.  Unfortunately, for the generic anisotropic model, the three
identities cannot be satisfied.  However, once we have identified the
point where the proof breaks down in the general case, we can look for
special anisotropic models that would still make the argument go through.
In this way, we find that the XXZ model
$$H = \sumL
[\s^x_j\s^x_{j+1}+\s^y_j\s^y_{j+1}+\Delta\s^z_j\s^z_{j+1}],\eqlabel\xxzham$$
with $\Delta=-1$ satisfies all the conditions.  Its conservation laws have then
a Catalan tree pattern.  This solution appears not to have a $su(N)$ extension.


\newsec{Algebraic framework}

In this section we formulate quantum ``spin" chains in a general algebraic
framework and define the sequence of charges with Catalan tree
pattern.
Recall first  a few basic definitions
[\ref{Schafer, R.D.,
{An Introduction to Nonassociative Algebras}. New York: Academic Press
1966}\refname\Schaf].
An algebra
over a field $\bf K$ is a vector space $A$ over $\bf K$ equipped
with a product (a mapping $A \times A \to A$) satisfying
the properties
$$\eqalign{
(\alpha+\beta) x=& \alpha x+ \beta x,\cr
\alpha (x+y) =& \alpha x+ \alpha y,\cr
\alpha (xy) =& (\alpha x)y=x( \alpha y)
} \eq$$
for all $\alpha,\beta$ in $\bf K$ and $x,y$ in $A$.
An algebra is associative
if the product is associative, i.e.
$$ (xy)z=x(yz) \eq$$
for all $x,y,z$ in $A$.
An algebra is alternative if
$$x(xy)=x^2 y {\text {and}}  (yx)x=y x^2 \eq$$
for all $x, y\in A$.
It follows immediately that any associative algebra is alternative.

We will consider models defined on a lattice
$\Lambda$
which may be either infinite ($\Lambda=Z$)
 or finite; in the latter case,
periodic boundary conditions are assumed ($\Lambda=\{1,\dots, N\}$, with
$N+1\equiv 1$).
Let $A$ be a finite-dimensional
(possibly nonassociative) algebra
with unity over a field $\bf K$.
We denote by $\calA$ the tensor product algebra
$${\cal A}=\bigotimes_{i \in \Lambda} A. \eq$$

Let us recall now some definitions from [\GMb].
Let $\calB^{(n)}(\L)$ be the set of all $n$-element sequences of points in
$\L$.
We define $n$-clusters as ordered elements of $\calB^{(n)}$, i.e.
an $n$-cluster is a sequence of
$n$ lattice sites ${\calC=\{{i_1}, ..., {i_n}\}}$,
with $i_1<i_2<...<i_n$.
We denote the set of $n$-clusters as $\calC^{(n)}(\L)$.
Let
$$\calB(\L)=\bigcup_{n=1}^{|\L|} \calB^{(n)}(\L) , \quad
\calC(\L)=\bigcup_{n=1}^{|\L|} \calC^{(n)}(\L),
$$
where $|\L|$ is the number of sites in $\Lambda$.
The sequences
in $\calB(\L)$ which are not in $\calC(\L)$ will be called
disordered clusters. For any $n$-cluster
we define the number of its holes $n_h$, i.e. the number of sites
in between $i_1$ to $i_n$  that are not included in $\calC$:
 $n_h(\calC)=i_n-i_1+1-n$. Let
$$\calC^{(n,k)}(\L)=\{ \calC\in \calC^{(n)} | n_h(\calC)=k\} \eq$$
be the set of all $n$-clusters of $\Lambda$
with $k$ holes.

Let $h: \calB(\L) \to \calA$ be some function and let
$H=\sum_{\calC\in \calB(\L)}h(C)$.
A triple $(A, \L, H)$ defines then a general
quantum chain with a hamiltonian $H$ on a lattice $\L$.

 We also recall the recurrence relation defining generalized Catalan numbers
$\alpha_{k,\ell}$:
$$\alpha_{k,\ell}=\alpha_{k-1,\ell-1} +\alpha_{k,\ell+1},\eq$$
with the understanding that  $\alpha_{k,\ell}=0$ if $\ell>k$.

Consider now an arbitrary function $\ft: \calB(\L) \to \calA$.
We will use the notation $\ft_n$ to denote a restriction of $\ft$ to
$n$-clusters:
$\ft_n=\ft|_{\calC^{(n)}(\L)}$.
We next define an elementary  operation (which is a derivation if
$A$ is associative)
$\delta_{ij}$ by
$$\delta_{ij} X \equiv  [\ft_2(i,j),X ] \eq$$
for $X$ in $\calA$.
To any such function $\ft$ we can associate a sequence $\{H_n(\ft)\}$
($n=2, \dots,|\L|$)
of elements of $\calA$, defined by
$$ H_n= F_{n,0}+ \sum _{k=1}^{[n/2]-1} \sum_{\ell=1}^{k} \alpha_{k,\ell}
F_{n-2k,\ell}, \eqlabel\catahn$$ where
the square bracket indicates the integer part and
$$ F_{n,k}=\sum_{\calC \in \calC^{(n,k)}} \ft(\calC),\eqlabel\fff$$
We will call (\catahn) a Catalan tree sequence
corresponding to the function $\ft$,
since each element of this sequence can be represented in terms
of a simple tree, known as Catalan tree [\GMa-\GMb].\foot{If
the characteristic of $\bf K$ is not zero, (\catahn) should be understood
as
$$ H_n= F_{n,0}+ \sum _{k=1}^{[n/2]-1} \sum_{\ell=1}^{k}
\sum_{m=1}^{\alpha_{k,\ell}}
F_{n-2k,\ell}$$.
}
We also define $$ \delta_{H_2}X=\sumL \delta_{jj+1} X
=[H_2, X]\eq$$
for $X$ in $\calA$.
Note that the elements of the Catalan tree sequence are
given by local expressions,
in the sense that terms involving a certain set of sites vanish
when the distances between the sites become sufficiently large.
A Catalan tree sequence will be called conserved if
$$ [H_2,H_n]=\sumL \delta_{jj+1}H_n=0.\eq$$

\newsec{ Sufficient  conditions for the conservation of the family
$\bf \{ H_n \}$ }

We first formulate the following technical:

\noindent {\bf Lemma}

{\it Suppose that there exist a function $\ft: \calB(\L) \to \calA$ such that
the
operation $\delta_{ij}$ has
the following properties:

\n (i) for any $n$-cluster $\calC=\{i_1,\dots, i_n\}$
$$\eqalign{
\delta_{jj+1}\ft(\calC)
=& - \aone \ft ({i_1-1} ,i_1, \dots,i_n) \text{if} j=i_1-1, \cr
=& \aone \ft (i_1,\dots,i_n,{i_n+1})\text{if} j=i_n, \cr
= & \aone \ft( {i_1},{i_1}+1, i_2,\dots,i_n )
\text{if} j=i_1\not=i_2-1,\cr
=&- \aone \ft(i_1,\dots,{i_n-1}, {i_n})\text{if}
j+1=i_{n}\not=i_{n-1}+1,\cr
=& 2 \atwo [ \ft( i_2,i_3,\dots,i_n)
- \ft(i_1,i_3,\dots,i_n )] \text{if} j=i_1=i_2-1, \cr
=& 2 \atwo [ \ft( i_1,\dots,i_{n-2},i_n)
-  \ft( i_1,\dots,i_{n-2},i_{n-1})]
\cr &\quad \text{if} j=i_{n-1}=i_n-1, \cr
=&\athree
 [ \ft(i_1,\dots,i_{k-1},i_{k+1},\dots,i_n)-
\ft(i_1,\dots, i_k,i_{k+2},\dots,i_n)]
\cr &\quad\quad
 \text{if} j=i_k=i_{k+1}-1 , k\not=1,n-1.
}\eqlabel\comlik$$

\n (ii) if $j=i_k$ and $i_{k+1}\ne i_{k}+1$
$$ \delta_{jj+1}[\ft(i_1,\dots,i_{k-1},i_k, i_{k+1},\dots,i_n)+
\ft(i_1,\dots,i_{k-1},i_k+1, i_{k+1},\dots,i_n)]=0
\eq$$
Then we have:

(a)
The Catalan tree sequence $H_n$ corresponding to $\ft$
is conserved
$$ [H_2(\ft),H_n(\ft)]=0\eq $$
iff
$$\aone+\athree=2 \atwo.\eq$$

(b) Moreover, if $A$ is  alternative
and if $\aone=\atwo=\athree$, the
Catalan tree sequence $\{H_n(\ft)\}$ forms a mutually commuting
associative family:
$$[H_n(\ft), H_m(\ft)]=0 \text{for all} n,m\ge 2. \eqlabel\hnmut$$

}

\n{\it  Proof}:  Part (a) of the
lemma can be proved by a direct calculation,
which is essentially a straightforward extension of the
proof presented in [\GMa-\GMb] for the case when $A=M_2(\bf C)$
(the algebra of complex $2\times 2$ matrices),
and which will not be repeated here. The proof of  part (b)
makes use of the properties of the operation:
$$\delta_B=\sumL \; j ~ \delta_{jj+1}=[ B, ~],\eqlabel\bdef$$
where
$$ B=\sumL \; j\,  \ft(j,j+1). \eq$$
Suppose first that the chain is infinite $|\L|=\infty$.
If $\aone=\atwo=\athree$ then $\delta_B$
generates recursively all the $H_{n>2}$ starting from $H_2$.
This can be seen from the fact that
$ \delta_B H_n$ is a linear combination of elements $H_{n+1-2k}$,
$k=0,\dots, [n/2]$. More exactly, we have:
$${ H_{n+1}={1\over{ (n-1)}} [B, H_n] +R_n,}\eqlabel\bact$$
where $R_n$ is a linear combination of the charges $H_{m<n}$.
In the associative case, we can use (\bact) in
an inductive argument
to prove  (\hnmut).  We outline this argument below
(cf. [\GMb], section 4.2.(i)).
Assuming that $[H_n, H_m]=0$ for all $n,m<n_0$, we prove that
$[H_{n_0+1}, H_k]=0$, for $k<n_0$. For $k=2$ this holds by
construction.  For $k=3$, the commutativity of
$H_{n_0+1}$ and $H_3$ can be established using the Jacobi identity and the fact
that $[H_{n_0+2}, H_2]=0$. Similarly, one may successively show that
$[H_{n_0+1}, H_{k>3}]=0$ using the Jacobi identity and the
relations $[H_{n_0+\ell}, H_2]=0$ for $\ell=1,\dots,k-1$.

If the algebra $A$ (and hence $\calA$) is
nonassociative, the Jacobi identity does not hold. However,
when $\calA$ is alternative we have
[\Schaf]
$$ [[ x,y],z]+[[y,z],x]+[[z,x],y]=6 (x,y,z), \eqlabel\altJac$$
for all $x,y,z \in \calA$,
where
$$(x,y,z)=(xy)z -x(yz)\eq$$
is the associator.
Moreover,  the Artin's theorem [\Schaf]
shows that the subalgebra of an alternative algebra generated by any two
elements is associative. Since  $H_{n\ge2}$  belong to the
subalgebra generated by $B$ and $H_2$, we have then
$$(B, H_n, H_m)= (H_n, H_m, H_k)=0 \eq$$
for all $n,m,k\ge 2$. In consequence, the Jacobi identity holds
for the commutators of the type $[B, [H_n, H_m]]$ and the
inductive proof of (\hnmut)  goes through.
Although strictly speaking the action of $\delta_B$
is incompatible with periodic boundary conditions,
the formula (\bact) remains valid for finite chains, if coefficients in
(\bdef) are understood modulo $|\L|$.
In consequence, (\hnmut) holds also for finite chains.
$\diamond $

In physical terms, part (a) of the above lemma can be
interpreted as giving a sufficient condition for the existence of
a family of conservation laws in the chain with the hamiltonian $H_2$.
Similarly, part (b) gives a sufficient condition for the integrability of
such chain. However, the lemma does not indicate how to
construct a function $\ft$ with the desired properties.
In the following, this lemma will be used as
a tool for verifying explicit constructions of $\ft$.

Next we present a theorem identifying sufficient conditions
for the existence of $\ft$ with the properties (i)-(ii),
when the multiplication in $A$ satisfies
certain constraints.
Clearly any such function can be determined from its restrictions to
$\calC^{(n)}(\L)$ (denoted $\ft_n$)
for each $n$, which we will construct below.

Let $u^{\alpha}$ ($\alpha=0,...,d-1$) be a basis in $A$. The product in
$A$ is completely specified by the multiplication table
$a^{\alpha\beta}_{\gamma}$, defined
by
$$ u^{\alpha}u^{\beta}=a^{\alpha\beta}_{\gamma} u^{\gamma}.\eq$$
Identifying  $u^0$ with the identity, $u^0\equiv 1$, we have then
$$a^{\alpha 0}_{\gamma}=a^{ 0 \alpha}_{\gamma}=\delta^{\alpha}_\gamma,
\quad a^{00}_0=1.\eq$$
For an algebra with unity we have then $(d-1)^3$ elements of the field
$\bf K$ uniquely determining the algebra. We decompose the structure constants
$a^{\alpha\beta}_{\gamma}$ into a symmetric and an antisymmetric part:
$$ \eqalign{
[u^a, u^b]= &c^{abg} u^g + \eta^{ab} u^0,\cr
\{u^a, u^b\}= &d^{abg} u^g + \theta^{ab} u^0},\eq$$
(the Latin indices go from 1 to $d-1$, the Greek ones from 0 to $d-1$).
The antisymmetric part $c^{abg}$ can be used to define
a ``vector product"
$$({\bf u}_{i}\times {\bf u}_{j})^g =
\beta^{-1} c^{abg}u^a_iu^b_j,\eqlabel\vecdef$$
where the vector ${\bf u}_i$ is defined in terms of its components $u^a_i$,
and $\beta$ is a nonzero constant.
We also define dot product
in a natural way, i.e.
$${\bf u}_i\cdot {\bf u}_j= \delta^{ab}u^a_i  u^b_j\eq$$
Assigning on each site of a cluster ${\calC=\{{i_1}, ..., {i_n}\}}$ a
basis element, we then construct $n$-linear polynomials
$$f_n(\calC)={\bf v}_{n-1}\mult {\bf u}_{i_n},\eqlabel\fndef$$
where the vector $\V_{n-1}$ is
obtained from the nested vector product of the basis vectors $u^a$ at
the first $n-1$ sites of the cluster:
$$\eqalign{
\V_1 =&\;{\bf u}_{i_1},\cr
\V_2=&({\bf u}_{i_1}\times {\bf u}_{i_2}),\cr
\V_3=&(({\bf u}_{i_1}\times {\bf u}_{i_2})\times{\bf u}_{i_3}),\cr
...&\cr
\V_{m}=&(\V_{m-1} \times {\bf u}_{i_m}).\cr}\eq$$
In other words,
$$ f_n(\calC)= (\dots({\bf u}_{i_1}\times  {\bf u}_{i_2})\dots  )\times
{\bf u}_{i_{n-1}})\cdot {\bf u}_{i_n}. \eqlabel\fnequ$$
If the tensor $c^{abe}$ is cyclic\foot{Note
that the requirement of cyclicity  for $c^{abc}$ implies that it
is a completely antisymmetric tensor.},
the polynomials $f_n$ have the property that
the dot product
can be placed at an arbitrary position, provided that parentheses to its left
(right) are nested toward the left (right), e.g:
$$ f_n(\calC)= ( {\bf u}_{i_1}\cdot ( {\bf u}_{i_2} \times ( {\bf u}_{i_3}\dots
\times ({\bf u}_{i_{n-1}} \times {\bf u}_{i_n})\dots). \eq$$
We form next the sequence $H_n(f)$. The first few elements are
$$\eqalign{
H_2=&\sumL \u_j \cdot \u_{j+1},\cr
H_3=&\sumL (\u_j \times \u_{j+1}) \cdot \u_{j+2},\cr
H_4=&\sumL [((\u_j\times \u_{j+1})\times \u_{j+2})\cdot \u_{j+3}
 + \u_j \cdot \u_{j+2}],\cr
H_5=&\sumL [(((\u_j\times \u_{j+1})\times \u_{j+2})\times \u_{j+3})\cdot
\u_{j+4}
 \cr &+ (\u_j \times \u_{j+2}) \cdot \u_{j+4}
 + (\u_j \times \u_{j+3}) \cdot \u_{j+4}
].
}\eq$$
We have then the  following

\n {\bf Theorem}

{\it
Suppose  that the structure constants of an algebra with unity
$A$ satisfy the
following relations:

\n (a)  $c$ is cyclic:
$$ c^{abg}=c^{bga}=c^{gab} \eq $$
(b) there exist constants $b_2,\;b_3$ such that
$$ \eqalign{
c^{abc}(\theta^{sa} c^{sbl}+\eta^{sa}d^{sbl}) =& \btwo \delta^{lc},\cr
c^{lap}c^{pbr}(\theta^{sa} c^{sbm}+\eta^{sa}d^{sbm}) =& \bthree c^{lmr},\cr
}\eq$$

\n (c) for all $c,l,r,m$:
$$\eqalign{
c^{abc}(\eta^{sa} \theta^{sb}+\theta^{sa}\eta^{sb}) =&
c^{abc}(c^{sal} d^{sbr}+d^{sal}c^{sbr}) =0, \cr
c^{lap}c^{pbr}(\eta^{sa}\theta^{sb}+\theta^{sa}\eta^{sb})=&
c^{lap}c^{pbr}(c^{sam}d^{sbc}+d^{sam}c^{sbc}) =0 .
}\eq$$
Then
if $$2 \beta^2+\bthree= \btwo\eq$$
the  Catalan tree sequence
$\{H_n(f)\}$ is conserved:
$$[H_n(f), H_2(f)]=0.\eq$$
}

\n {\it Proof:}  The theorem is proven by  a direct calculation verifying
that
when the constraints (a)-(c) hold,
the function $f$ defined  in (\fndef)
satisfies the conditions (i) and (ii) of the preceding lemma
with
$$\aone=\beta, \quad 2 \atwo = \btwo/(2 \beta) , \quad
\athree=\bthree/(2 \beta)\eq$$
(recall
that $\beta$ is the constant in the definition (\vecdef)).
 $\,\diamond $

The assumptions
of the theorem
may not seem to be very transparent. However,
the conditions (a)-(c) above
may be equivalently
rewritten in terms of the following three simple
identities.
Let ${\bf L}$
and ${\bf R}$ be
vectors build out of the algebra generators, but involving only sites
on the left of $i$ and on the right of $i+1$
respectively. The identities are:
$$[{\bf u}_{i}\cdot {\bf u}_{i+1}, {\bf u}_{i+1}\cdot {\bf R}]=
-\aone ({\bf u}_{i}\times{\bf u}_{i+1}
)\cdot {\bf R}, \eqlabel\idone$$
$$ [{\bf u}_i\cdot {\bf u}_{i+1},
({\bf u}_i\times {\bf u}_{i+1})\cdot {\bf R}]=
2 \atwo \{( {\bf u}_{i+1}\cdot {\bf R}) -
 ({\bf u}_{i}\cdot {\bf R})\}, \eqlabel\idtwo$$
$$[{\bf u}_i\cdot {\bf u}_{i+1},
 (({\bf L}\times {\bf u}_i)\times {\bf u}_{i+1})
\cdot {\bf R}]=
\athree \{( {\bf L}\times {\bf u}_{i+1})\cdot {\bf R}
- ( {\bf L}\times {\bf u}_{i})\cdot {\bf R}\}.
\eqlabel\idthree$$
These identities are equivalent to the conditions (a)-(c) above if
$\aone=\beta$, $2 \atwo = \btwo/(2 \beta)$, and $\athree=\bthree/(2 \beta)$.

Moreover, a simple calculation shows that if the assumptions of the theorem
hold,
$$H_1^a=\sumL u^a_j \eqlabel\Hone$$
commutes with $H_2$.

Note that the construction of the
Catalan tree sequence makes use of an explicit choice of
basis in the algebra $A$. In particular,
the assumptions of the Theorem in section 2 requires cyclic symmetry of
$c^{abc}$: this property is basis dependent.

The Catalan tree sequence of charges for the  $SU(N)$-invariant
models corresponds to
$A=M_N({\bf C})$.
A convenient basis for
$M_N({\bf C})$ is provided by the unit matrix $\bf I$ and the set
of $su(N)$ Gell-Mann matrices $t^a$,
($a=1,\dots, N^2-1$)
satisfying
$$\eqalign{ [t^a,t^b]=&2  i f^{abc} t^c, \cr
t^a t^b +t^b t^a=&4( \delta_{ab}/M) {\bf I}+2\hat d^{abc}t^c,}\eq$$
where $f^{abc}$ are the structure constants of $su(N)$, and
$\hat d^{abc}$ is a completely symmetric tensor, nontrivial for all $N>2$.
Choosing the $u^\alpha$ basis as $u^0={\bf I}$,
and
$u^a=t^a$, ($a=1, \dots, N^2-1$) we have then
$$c^{abg}=2i f^{abg}, \quad \theta^{ab}=(4/N) \delta^{ab},\quad d^{abc}=
2i\hat d^{abc}.\eq$$
With the choice of $\beta=2i$
for the constant in the vector product,
we obtain the identities
(\idone-\idthree) with $\aone=\atwo=\athree=2 i$.

In the subsequent sections, we will present two
other examples of systems with a Catalan tree sequence of charges.

\newsec{The isotropic octonionic chain}

In this section we consider the algebra of octonions $\bf O$, a nonassociative
alternative division algebra over $\bf R$.
Let $\{e^0=1, e^a\}$ ($a=1,\dots,7$) be a basis in ${\bf O}$,
with the
octonionic multiplication  defined by
$$e^a e^b=-\delta^{ab} + f^{abc} e^c, \eq$$
where $f^{abc}$ are the Cayley structure constants. There are 480
different possible choices for the multiplication table; one
aesthetically appealing choice,  which we will adopt here, is:
$f^{abc}=1 $ for the cycles $$(abc)=(123), \, (246),\,
(435), \,(367)\, (651),\,(572),\, (714).\eq$$
The Cayley structure constants obey the identity
[\ref{D\"undarer, R.,  G\"ursey, F., Tze, C.-H.: Generalized vector
products, duality and octonionic identities in $D=8$
geometry. {J. Math. Phys.} {\bf 25}, 1496-1506 (1984)}]:
$$f^{abc}f^{gec}=\delta^{ag}\delta^{be}-\delta^{ae}\delta^{bg}+
\phi^{abge}, \eqlabel\idoct$$
where $\phi^{abge}$ is a completely antisymmetric tensor defining
the octonionic associator: $$
(e^a,e^b,e^g)=(e^a e^b) e^g -e^a (e^b e^g)=2 \phi^{abge} e^e\eq $$
(with the understanding that $a, b, g$ above are all different).
Tensors $\phi$ and $f$ are dual to each-other:
$$ \phi^{abcd}=-{1\over {3! }} \epsilon^{abcdegh}f^{egh}. \eq$$
The identity (\idoct) implies that
$$ f^{abc}f^{gbc}=6\, \delta^{ag}.\eq$$
We have also
$$ f^{abc}f^{cge}f^{epa}=3 f^{bgp}. \eq$$
In the basis $u^a=e^a$, $u^0=e^0$
we have then $$c^{abc}=2 f^{abc},\quad \theta^{ab}=-2 \delta^{ab}, \quad
\eta^{ab}=d^{abc}=0. \eq$$
We define the vector product as
$$({\bf e}_j\times {\bf e}_k)^c= \gamma^{-1} f^{abc}e^a_j e^b_k, \eq$$
where $\gamma$ is a constant.
Using the identities for the products of the Cayley structure
constants, we find that the identities (\idone-\idthree) are
satisfied when
$$\aone=2 \gamma, \quad 2 \atwo=-12 /\gamma, \quad
\athree=6/\gamma.\eq$$
Therefore the condition
$\aone+\athree=2 \atwo$ is satisfied for the
choice of $\gamma= \pm 3i$.
The value of this constant turns out to be complex,
which may be accounted for by
extending the octonionic algebra to $$A={\bf C}\otimes {\bf
O}\eqlabel\octcomp$$
This complexified octonionic algebra is still an alternative algebra, but
no longer a division algebra, as it has
nontrivial zero divisors.

By the theorem in section 3
the sequence $H_n(f)$
(defined by (\catahn) with $f$ defined in (\fnequ))
is conserved
for $\gamma =\pm 3i$.
In other words, the octonionic model with the hamiltonian
$$ H_2=\sumL e^a_{j} e^a_{j+1}\eq $$
has   $|\L|-2$
conservation laws $H_{n>2}$. The first few are:
$$\eqalign{
 H_3=&\sumL \gamma^{-1} f^{abc} e^a_j e^b_{j+1} e^c_{j+2}, \cr
H_4=&\sumL [\gamma^{-2} f^{abs} f^{scd} e^a_j e^b_{j+1} e^c_{j+2} e^d_{j+3}+
e^a_j e^a_{j+2}],\cr
H_5=&\sumL \gamma^{-1}[\gamma^{-2} f^{abs} f^{scm}f^{mdh}
 e^a_j e^b_{j+1} e^c_{j+2} e^d_{j+3} e^h_{j+4}\cr &+
f^{abc}(e^a_j e^b_{j+2}e^c_{j+3}
+e^a_j e^b_{j+1}e^c_{j+3})
] .} \eq$$
This model appears as the octonionic generalization of the Heisenberg
(XXX) model
$$ H=\sumL \s^a_j  \s^a_{j+1}, \eq $$
where $\s^a$, $a=x,y,z$  are the Pauli sigma matrices,
or equivalently imaginary unit quaternions.

We can see that
both choices for sign of $\gamma$ lead to equivalent results
(under a change $\gamma\to -\gamma$ the charges $H_n$ of even order are
invariant, while the charges of odd orders change signs).
Observe also that a trivial redefinition $H_{2k+1} \to \gamma H_{2k+1}$
gives a system of conserved charges with real coefficients,
thus yielding conservation laws in $\bigotimes_{j\in\L}{\bf O}$
(without the complexification (\octcomp)).

Note that for $\gamma=\pm 3i$ the condition $\aone=\atwo=\athree$
is not satisfied, and
 hence part (b) of the lemma in section 3 cannot be used.
Mutual commutativity of the $H_n$'s (which is needed to classify
the octonionic chain as an integrable system)
 must be  checked
independently. Surprisingly, higher order charges do not commute.\foot{Note
that (up to a constant) $[B, H_2]=H_3$,
but $[B,H_3]$ is not a conserved quantity. Thus there is no
boost construction for the charges in this case.}
In particular, for $|\L|\ge 6$, $[H_3, H_4]$ does not vanish.\foot{
For  $|\L|=4,5$ $[H_3, H_4]=0$. }

\newsec{The XXZ model with $\Delta = -1$}

Can the pattern of conserved charges in an anisotropic chain be described in
terms
of a Catalan tree?
To investigate this problem, we concentrate on the XYZ model,
and analyze possible modifications of the three identities
(\idone)-(\idthree) to take the anisotropy into account.
We will see
that this leads in a natural way to a construction of a
function $\ft$ satisfying the
requirements of the lemma in section 3, which is a kind of deformation of
the polynomials $f$.

The XYZ hamiltonian to be considered is
$$H = \sumL
[\a\s^x_j\s^x_{j+1}+\b\s^y_j\s^y_{j+1}+\c\s^z_j\s^z_{j+1}],\eqlabel\xyzham$$
where
$\a,\b,\c$ are constants, and $\s_j^a$'s  are the Pauli matrices.  We rewrite
it
in the form
$$H = \sumL
\sh^a_j\sh^a_{j+1},\eq$$
where
$$\sh^a_j =\sqrt{\la_a}~\s^a_j.\eq$$
At first sight, a natural guess for the modified form of the  $f$ polynomials
is to
replace all spin factors by their hat versions (which ensures
that $H_2$ coincides with the XYZ hamiltonian).
To see whether this is appropriate,
we first consider the commutator of
$ {\bf \Sh}_{i}\cdot {\bf \Sh}_{i+1}$ with $ {\bf \Sh}_{i+1}\cdot {\bf R}$,
where ${\bf R}$
involves only  spin
variables at sites  to the right of $i+1$:\foot{
In the rest of this section summation over indices repeated two or more times
is implied, unless indicated otherwise.}
$$[\sh^a_i\sh^a_{i+1}, \sh^b_{i+1} R^b] = 2i\sqrt{{\la_a\la_b\over
\la_c}}~\ep_{abc}~
\sh^a_i\sh^c_{i+1} R^b\eq$$
The goal is to express commutators of $ {\bf \Sh}_{i}\cdot {\bf \Sh}_{i+1}$
with
the
polynomials $f_n$ in terms of higher or lower order but similar polynomials.
Due the to extra factor  $\sqrt{\la_a\la_b/ \la_c}$, the right hand side
of the above equation cannot be rewritten in the form
 $({\bf
\Sh}_{i}\times {\bf \Sh}_{i+1})\cdot {\bf R}$.  However,
we can rescale the spin variable at site $i+1$ by
$$\st^c_{i+1}=
{\sqrt{ {\a\b\c} \over {\la_c} }}~ \s^c_{i+1}\eq$$
In this way, we find
$$[{\bf \Sh}_{i}\cdot {\bf \Sh}_{i+1}, {\bf \Sh}_{i+1}\cdot {\bf R}] = -2i({\bf
\Sh}_{i}\times {\bf \St}_{i+1})\cdot {\bf R}\eq$$
This suggests to replace in the expression for the $f$'s, the spins at the
border of the cluster by their hat versions and all the other ones by their
tilde
versions. But this is not enough.  As the next calculation will show, an extra
multiplying factor, depending on the position of the holes, has to be
introduced.  Consider:
$$ [{\bf \Sh}_{i}\cdot {\bf \Sh}_{i+1},({\bf \Sh}_{i}\times{\bf
\St}_{i+1})\cdot
{\bf R}] = \la_a\la_c\sqrt{\la_e}~\ep_{cde}[\s^a_i\s^a_{i+1},
\s^c_i\s^d_{i+1}] R^e.\eqlabel\comone$$
Using
$$2[\s^a_i\s^a_{i+1},
\s^c_i\s^d_{i+1}] = [\s^a_i,
\s^c_i]\{\s^a_{i+1},
\s^d_{i+1}\} +\{\s^a_i,
\s^c_i\}[\s^a_{i+1},
\s^d_{i+1}] ,\eq$$
(\comone) is found to be
$$2i\la_a\la_c\sqrt{\la_e}~\ep_{cde}~ (\delta_{ad}\ep_{acf}~\s^f_i +
\delta_{ac}\ep_{adf}~\s^f_{i+1})\,R^e.\eqlabel\secid$$
To investigate the effect of holes, we concentrate on the first term.
We cannot use directly  the identity
$$\ep_{cae}\ep_{caf} = 2\delta_{ef},\eq$$
to simplify this term,
because the prefactor depends on the indices $a$ and $c$, over
which we want to sum.
But the cure is simple: one just notices that
$$ \la_a\la_c~\ep_{cae} = {\a\b\c\over \la_e}~\ep_{cae}\eqlabel\transfo$$
(no summation).
With this transformation, the troublesome prefactor is eliminated and
the first term in (\secid) becomes
$$-4i{\a\b\c\over \la_e}~\s^e_iR^e. \eq$$
This shows that to
account for the presence of a hole, here at site $i+1$, an extra
factor
$\a\b\c/ \la_e$ has to be introduced, where the index $e$ stands for the
overall
component of the vector product to the right (or, equivalently to the left)
 of the hole. Recall
that (as a consequence of the cyclicity of $c^{abg}$)
in the $f$-type polynomials, the dot (scalar
product) can be placed anywhere, provided that all vector products at its left
are nested toward the left and all those at its right are nested toward the
right.
In the XYZ context, it is nothing
but a retranscription of the familiar vector identity
$$({\bf A}\times {\bf B})\cdot{\bf C} = {\bf A}\cdot({\bf B}\times {\bf
C}).\eq$$
The component of the vector product say to the left of a hole is easily read
off when the dot
product is placed at the left of the hole site.

We now complete the evaluation of (\secid), whose second term reads
$$ 2i\la_a^2\sqrt{\la_e}~ \ep_{ade}\ep_{adf}~\s^f_{i+1}R^e.\eq$$
But it presents a serious difficulty: a prefactor depending upon the index $a$
(or the index $d$, after a transformation of the type (\transfo)) cannot be
avoided, so that the result cannot be written in terms of a scalar product.
The
only way to proceed is to assume that all $\la_a^2$ are the same.  In that
case, without loss of generality, we can let $\la_a^2=1$.  For this
second term, we get then
$$4i{\bf \Sh}_{i+1}\cdot {\bf R}.\eq$$

Let us summarize the situation at this point in the anisotropic case.  We have
found that for the general XYZ model, we cannot prove the second
of the three identities, unless  $\la_a^2$ is independent of $a$.  With
$\la_a^2=1$,
this gives two independent solutions (up to  trivial relabelings of the
space
directions), namely the XXX isotropic model, where $\a=\b=\c=1$, and the XXZ
model
with $\a=\b=1$  and $\c\equiv\Delta=-1$.  This second solution, denoted ${\rm
XXZ}_{-1}$ from now on, is a new candidate for a model with a Catalan tree
pattern.
Moreover, by studying these first two identities, we have seen how the
polynomials $f$ need to be modified in the anisotropic case:
$$f_n\ra \ft_n = g_{\{{\rm holes}\}} ((...((\Sh_{i_1}\times\St_{i_2})\times
\St_{i_3})...)\times\St_{i_{n-1}})\cdot \Sh_{i_n},\eq$$
where $g$ is the product of the factors $\a\b\c/ \la_{e_j}$, one for each
hole ($j=1,...,k$), where $e_j$ is the overall group index at the left of the
hole
at site $j$.  When
$\la_a^2=1$, $g$ is just a sign. (Note that we have just
rederived, in a different
way, the rules for constructing these
anisotropic polynomials obtained in [\GMb] from the boost construction of the
charges.)

Consider now the third identity in the context of the ${\rm XXZ}_{-1}$ model.
A simple calculation yields
$$[{\bf \Sh}_{i}\cdot {\bf \Sh}_{i+1}, (({\bf L}\times{\bf \St}_{i})\times {\bf
\St}_{i+1})\cdot {\bf R}]=2i \{g_{\{i\}}({\bf L}\times{\bf \St}_{i+1})\cdot
{\bf R}-
g_{\{i+1\}}({\bf L}\times{\bf \St}_{i})\cdot {\bf R}\}.\eq$$
Hence, with appropriate correction factors and rescalings, this identity is
also
satisfied.

In the isotropic
case, the three identities ensure the validity of all the steps in the proof of
the
commutativity of
$H_n$ with the hamiltonian: in other words the function $f$ satisfies both
the conditions (i) and (ii) of the lemma in section 3.
But when these identities are
modified as above, only part (i) follows automatically. A key step
of the argument, part (ii),
must be reconsidered entirely.
In the language of [\GMb] this amounts to establishing
the cancellation of the terms corresponding to
disordered clusters.
Such terms have the form
$$({\bf L}\times({\bf \St}_{i}\times {\bf
\St}_{i+1}))\cdot {\bf R}\not=(({\bf L}\times{\bf \St}_{i})\times {\bf
\St}_{i+1})\cdot {\bf R},\eq$$
and they arise in the commutator of ${\bf \Sh}_{i}\cdot {\bf \Sh}_{i+1}$ with
$\ft_n(\calC)$ for clusters $\calC$ having a hole either at $i$ or $i+1$. In
the
appendix, we use the Jacobi identity to show that such terms can be rewritten
as a
sum of fully nested terms plus terms that cannot be nested.  When we sum
over clusters in (\catahn),
the latter part is shown to cancel for the general XYZ
case.   Hence, even though there is no Catalan tree pattern for the generic XYZ
model, the conserved charges can all be expressed in terms of sums over the
polynomials $\ft_n$, as proved in [\GMb].  For the particular
${\rm XXZ}_{-1}$ case, the first part also adds up to zero when summed over
appropriate clusters.
This ensures that the condition (ii) of the lemma holds and it is
enough to guarantee the Catalan tree pattern of conserved
charges.

To summarize: we have constructed a function $\ft$
satisfying the requirements of the lemma in section 3 with
$\aone=\atwo=\athree$. The elements of the
Catalan tree sequence corresponding to
$\ft$ are linear combinations of the charges obtained from
the logarithmic derivatives of the transfer matrix [\GMb], with $H_2$
coinciding with the XXZ$_{-1}$ hamiltonian.
By part (b) of the lemma, the elements of the Catalan tree
sequence $H_n(\ft)$ mutually commute (this of course is also
an immediate consequence of the transfer matrix formalism).

We present an example of a polynomial $\ft$.
For $\calC=\{1,2,5\}$, we have
$$\eqalign{
\ft_3(1,2,5) =& g_{\{3\}}g_{\{4\}}(\Sh_1\times \St_2)\cdot\Sh_5,\cr
=&({\a\b\c\over
\la_c})^2~\ep_{abc}\sh_1^a\st_2^b\sh_5^c,
\cr
=&({\a\b\c\over
\la_c})^2\sqrt{\la_a}\sqrt{{\a\b\c\over
\la_b}}\sqrt{\la_c}
{}~\ep_{abc}\s_1^a\s_2^b\s_5^c.
\cr}\eq$$
Specializing to the ${\rm XXZ}_{-1}$
model, we set
$$\eqalign{
\sh^x =&\s^x,\quad \sh^y =\s^y ,Ê\quad \sh^z =\sqrt{-1}\s^z,\cr
\st^x =&\sqrt{-1}\s^x,\quad \st^y =\sqrt{-1}\s^y ,Ê\quad \st^z
=\s^z.\cr}\eq$$
We give below the first few nontrivial charges of the XXZ$_{-1}$ model
read off from (\catahn). The first nontrivial charge beyond the hamiltonian is:
\eqn\hxyzt{
H_3=F_{3,0}=\sum_{j\in \Lambda} (\Sh_j\times \St_{j+1})\cdot \Sh_{j+2}.}
The four-spin charge is:\foot{We
have used the fact that ${ {\a\b\c} / {\lambda_a}}=-\lambda_a$ for the
XXZ$_{-1}$ model.}
\eqn\hxyzf{H_4=F_{4,0}+F_{2,1},}
where
$$\eqalign{
F_{4,0}=&\sum_{j\in\Lambda}
((\Sh_j\times \St_{j+1})\times\St_{j+2})\cdot \Sh_{j+3},\cr
F_{2,1}=&
-\sum_{j\in\Lambda} \lambda_a
\sh^a_j \sh^a_{j+2},
}\eq$$
The five-spin charge is:
\eqn\hxyzfif{H_5=F_{5,0}+F_{3,1},}
with
$$\eqalign{ F_{5,0}=&\sum_{j\in \Lambda}
(((\Sh_{j} \times \St_{j+1})\times
\St_{j+2})\times \St_{j+3})\cdot
\Sh_{j+4},\cr
F_{3,1}=&-
\sum_{j\in\Lambda}
\lambda_a \epsilon^{abc}(
\sh^a_j \st^b_{j+2}\sh^c_{j+3}
+\sh^b_j \st^c_{j+1}\sh^a_{j+3}).\cr
}\eq$$

In deriving the correction factors for the  polynomials $f$,
we have used in a
crucial way the fact that there are only three generators (in (\transfo) in
particular). This hints that the present analysis is most probably not
applicable
to algebras other than
$M_2({\bf C})$.
Indeed, we have not found anisotropic
$M_{N>2}({\bf C})$
models which would satisfy the generalized version of the three identities.

\newsec{Conclusions}

In this work we have established sufficient conditions for the existence of
a system of conserved charges with a Catalan tree pattern.
These conditions are formulated in the lemma and theorem of section 3.
The lemma is rather technical, and does not indicate how to
construct the sequence of charges from the underlying
algebraic structure.  The theorem gives
such  a construction, which is possible if the algebra obeys a number
of constraints.
These conditions can be viewed as being equivalent to the three equalities
(\idone)-(\idthree), which are
parametrized by three coefficients $\aone, \atwo, \athree$.
Our  construction gives a sequence of elements $\{H_n(f)\}$ such that
$[H_2(f), H_n(f)]=0$, provided that the coefficients $\aone, \atwo, \athree$
are related by the condition $\aone+\athree=2 \atwo$.

Because an overall rescaling of a function $\ft$ is irrelevant, and
due to the constraint $\aone+\athree=2 \atwo$ needed to
ensure the conservation of the Catalan tree sequence, this
construction gives a one-parameter family of systems. For
example, we can choose for the free parameter the ratio $\kappa=\atwo/\athree$.
For $A=M_N({\bf C})$  and for the XXZ$_{-1}$ case,
this ratio is $\kappa=1$; for the octonionic system described
in section 4,
$\kappa=-1$.
Each model in this family can be regarded as a quantum
chain with a hamiltonian $H_2$ and a sequence of
$|\L|-2$ conserved charges $\{H_{n>2}\}$ (we can also add
$H_1^a$ defined in (\Hone)). Since these systems possess an infinity of
conserved charges (when $|\L|=\infty$), they may be expected to be
integrable.
However,
the standard definition of quantum integrability requires
not only the existence of a family conserved charges, but their
mutual commutativity as well.
For $\aone=\atwo=\athree$, the commutativity
of the family $\{H_n\}$ (if the algebra is alternative)
can be established
by means of a recursive argument based on the properties of the
boost operator $B$.
This argument uses the Jacobi identity, or its generalization (\altJac),
and it breaks down for a nonalternative
algebra. In the general case
($\aone:\atwo:\athree\ne1$ or if $A$ is nonalternative)
mutual commutativity
of the conserved charges cannot be taken for granted, as the example
of the octonionic system shows.

A number of interesting problems remains to be studied. Are there other
models with the Catalan tree pattern of conserved charges?
In particular, is there such a model for every  value of $\kappa$?
When do the resulting charges mutually commute?
How does the presence of such
a special pattern of charges reflect itself in the Yang-Baxter
equations, the usual hallmark
of integrability?
Finally, the octonionic chain is interesting in its own right and
deserves further study.
\bigskip
\centerline{\bf Acknowledgment}
One of us (M.P. G.) wishes to thank  C.-H. Tze
for discussions concerning octonions.

\appendix{A}
{Cancellation of disordered clusters in the XXZ$_{-1}$ model}

We show in this appendix that in the summation over all clusters appearing
in (\catahn), terms
in $[{\bf \St}_{i}\cdot {\bf
\St}_{i+1},H_n]$ corresponding to disordered clusters
cancel two by two.  Such terms originate
from the polynomials of the form $g_{\{j\}}({\bf L}\times{\bf \St}_{j})\cdot
{\bf
R}$ with $j$  either $i$ or $i+1$. As usual, ${\bf L}$ and ${\bf R}$ are
assumed to commute with $\S_{i}$ and $\S_{i+1}$.  We will prove that the
non-nested
parts of
$$[{\bf \St}_{i}\cdot {\bf
\St}_{i+1}, \;
g_{\{i+1\}}({\bf L}\times{\bf \St}_{i}))\cdot {\bf R}+g_{\{i\}}({\bf
L}\times{\bf
\St}_{i+1}))\cdot {\bf R}]\eqlabel\nonorder$$
vanish.  This result holds true for the general XYZ model\foot{This
was stated in [\GMb] but without a detailed  proof.}.
The evaluation of the
first commutator  yields\foot{ As in section 5,
summation over all indices that appear twice or
more is implied, unless indicated otherwise.}
$$2i{\la_s(\a\b\c)^{3/2}\over
\la_c\sqrt{\la_b}}~\ep_{abc}\ep_{sbn}~L^a\s_i^n\s_{i+1}^sR^c.\eqlabel\ind$$
Using  (\transfo) and the Jacobi identity
$$\ep_{acb}\ep_{sbn} = \ep_{sab}\ep_{cbn}+\ep_{csb}\ep_{abn}\eq$$
(no summation) we get:
$$-2i{\a\b\c\la_s\sqrt{\la_s\la_n}\over
\la_c}(\ep_{sab}\ep_{cbn}+\ep_{csb}\ep_{abn})~L^a\s_i^n\s_{i+1}^sR^c.
\eqlabel\firstterm$$
The first part corresponds to a disordered term:
${\bf L}$ is first multiplied by the spin at site $i+1$ and the result
is multiplied by the one at site $i$. The second part is ordered and properly
nested.
The second commutator in (\nonorder) is evaluated in the same way, with the
result
$$-2i{\a\b\c\la_s\sqrt{\la_s\la_n}\over
\la_a}(\ep_{sab}\ep_{cbn}+\ep_{csb}\ep_{abn})~L^a\s_i^s\s_{i+1}^nR^c.
\eqlabel\secterm$$ Here the second term is not ordered.
The addition of the first term of (\firstterm) to the second term
of (\secterm) gives the total contribution of the unwanted terms,
corresponding to disordered clusters:
$$2i\a\b\c\sqrt{\la_s\la_n}\ep_{asb}\ep_{ncb}~L^a\s_i^n\s_{i+1}^sR^c
({\la_s\over
\la_c}- {\la_n\over
\la_a})\eq$$
Due to the presence of the antisymmetric tensors, we may
write
$${\la_s\over
\la_c}= ({\a\b\c\over \la_a\la_b})({\la_n\la_b\over \a\b\c})= {\la_n\over
\la_a}\eq$$
(no summation) which shows that the unwanted terms cancel. The sum of the
remaining terms  of
(\firstterm) and (\secterm) is
$$2i {\la_s\la_n\a\b\c\over \la_b}\ep_{asb}\ep_{ncb}~L^a\s_i^s\s_{i+1}^nR^c
({1\over
\la_a^2}- {1\over
\la_c^2}).\eq$$
This vanishes only if all $\la_a^2$ are identical, that is for the XXX or
the ${\rm XXZ}_{-1}$ model. In consequence, for these models (\nonorder) is
zero,
and thus the condition (ii) of the lemma in section 3 is satisfied. In
contrast,
for the
general XYZ model this condition does not hold.

\bigskip
\centerline{\sc{references}}
\immediate\closeout\refs \vskip 0.5cm
  \message{References}\input references

\end